%
\input phyzzx
\hfuzz 20pt
\font\mybb=msbm10 at 12pt

\def\Bbb#1{\hbox{\mybb#1}}

\def\bR{\Bbb {R}}

\def\bZ {\Bbb{Z}}

\def\bfomega{\omega\kern-7.0pt \omega}

\def\C{\mkern1mu\raise2.2pt\hbox{$\scriptscriptstyle|$}\mkern-7mu{\rm C}}


\def\hra{\hookrightarrow}

\REF\dai{J. Dai, R.G. Leigh, and J. Polchinski, 
{\sl New Connections between
String Theories}, Mod. Phys. Lett. {\bf A4} (1989) 2073.}
\REF\strom{A. Strominger, {\sl Open p-branes}, Phys. Lett. 
{\bf B383} 44; hep-th/9512059.}
\REF\pktone{P.K. Townsend, {\sl D-Branes from M-Branes}, Phys. Lett.
{\bf B373} (1996) 68, hep-th/9512062.}
\REF\gppkt{G. Papadopoulos and P.K. Townsend,  
{\sl Intersecting M-branes}, Phys. Lett.
 {\bf B380} (1996) 273; hep-th/9603087.}
\REF\pt{P.K. Townsend, {\sl Brane Surgery}, Nucl. 
Phys. Proc. Suppl {\bf 58} (1997) 163; hep-th/960921.}
\REF\pol{J. Polchinski, in \lq\lq Fields, Strings and Duality" 
(Tasi '96) Ed. C. Efthimiou and
B. Greene, World Scientific 1997; hep-th/9611050. }
\REF\wittenb{E. Witten, {\sl Five-brane Effective 
Action in M-theory} J. Geom. Phys. {\bf 22} (1997) 103; hep-th/9610234.}
\REF\gibbons{G.W. Gibbons, G.T. Horowitz and P.K. Townsend, 
{\sl Higher
Dimensional Resolution of Dilatonic Black Hole Singularities},
 Class. Quantum Grav. {\bf 12}
(1995) 297; hep-th/9410073.}
\REF\bott{R. Bott and L.W. Wu, {\sl Differential Forms in Algebraic
Topology}, Springer-Verlag (1982) Berlin.}
\REF\shrz{J. Schwarz, {\sl Lectures on Superstrings and M-theory Dualities},
Nucl. Phys. Proc. Suppl. {\bf B55} (1997) 1.}
\REF\maldaa{C. G. Callan and J.M. Maldacena, {\sl Brane Dynamics from
the Born-Infeld Action}, hep-th/9708147.}
\REF\bergpap{E. Bergshoeff, J.P. van der Schaar 
and G.Papadopoulos, {\sl Domain Walls on
the Brane}, Phys. Lett. {\bf B430} (1998) 63, hep-th/9801158. }
\REF\schwarzb{J.H. Schwartz, {\sl From Superstrings 
to M-theory}, hep-th/9807135.}
\REF\schwarz{J.H. Schwartz, {\sl The power of M-theory}, 
Phys. Lett. {\bf B367}
(1996) 97; hep-th/9510086. {\sl M-Theory extensions of T-duality}, 
hep-th/9601077.} 
\REF\itexas{I.V. Lavrinenko, H. L\"u, 
C.N. Pope and K.S. Stelle, {\sl Superdualities,
Brane Tensions and Massive IIA/IIB Duality},
hep-th/9903057.}
\REF\papastwo{G. Papadopoulos, {\sl T-Duality and the Worldvolume
Solitons Solitons of Five-Branes and KK-Monopoles},
 Phys. Lett. {\bf B434} (1998) 277, hep-th/9712162.}
\REF\tseytlinb{A.A. Tseytlin, {\sl Harmonic 
Superposition of M-Branes}, Nucl. Phys.{\bf B475}
(1996) 149.}
\REF\kastor{J.P. Gauntlett, D.A. Kastor and J. 
Traschen, {\sl Overlapping Branes in M-theory}
{\sl Nucl. Phys. } {\bf B478} (1996) 544.}
\REF\kal{K. Becker, M. Becker and A. Strominger, {\sl Fivebranes, Membranes
and non-perturbative string theory}, Nucl. Phys. {\bf B456} (1995), 130.
\break
E.Bergshoeff, R.Kallosh, T.Ortin and G.Papadopoulos,
{\sl Kappa-Symmetry, Supersymmetry and
Intersecting Branes}, Nucl.Phys. {\bf B502} (1997) 149.}
\REF\gp{G. Papadopoulos, {\sl A Brief Guide to p-Branes} in \lq\lq Recent
Developments in Theoretical Physics: STU-Dualities and Nonperturbative
Phenomena in Superstrings and Supergravity", Forch. 
Phys. {\bf 44} (1996) 573; hep-th/9604068.}
\REF\yang{H. Yang, {\sl Localized Intersecting Brane Solutions
of D=11 Supergravity}, hep-th/9902128.}
\REF\youm{D. Youm, {\sl Localized Intersecting BPS Branes}, hep-th/9902208.}
\REF\ansar{A. Fayyazuddin and D.J. Smith, {\sl Localized intersections
of M5-branes and four-dimensional superconformal field theories},
hep-th/9902210.}
\REF\loewy{A. Loewy, {\sl Semi Localized Brane Intersections in SUGRA},
hep-th/9903038.}
\REF\papgib{J.P. Gauntlett, G.W. Gibbons, 
G. Papadopoulos and P.K. Townsend, {\sl Hyper-K\"ahler
Manifolds and Multi-Intersecting Five Branes},  Nucl. Phys.
{\bf B500} (1997) 133; hep-th/9702202.}
\REF\peet{D. Marolf and A. Peet, {\sl Brane Baldness vs Superselection 
Sectors}, hep-th/9903213.}

\Pubnum{ \vbox{ \hbox{}\hbox{} } }
\pubtype{}
\date{May, 1999}
\titlepage
\title{Brane Surgery with Thom Classes}
\author{ G. Papadopoulos\foot{Address from 1st June 1999: 
Department of Mathematics, King's College London, 
Strand, London WC2R 2LS.}}
\address{DAMTP,\break Silver Street, \break  
University of Cambridge,\break Cambridge CB3
9EW}
\abstract {We propose a method to  investigate the
 conservation of brane charges at the intersection
of two or more branes using the Thom 
classes of their normal bundles.
In particular we find a relation 
between the charge of the branes
involved in the configuration and the 
charge of the defects on the
branes due to the intersection. We also 
explore the applications of our method for  various 
brane intersections in type II strings and M-theory.
}

\endpage
\pagenumber=2



\chapter{ Introduction}

One of the most remarkable properties of
branes is that they can  end or 
intersect with others [\dai,\strom,\pktone,\gppkt].
The arguments that lead us to 
believe that this is the case
are either based on brane charge 
conservation [\strom,\pt] or the analysis
of the couplings of brane 
effective theories. In same cases,
this can also be achieved by 
direct  construction like in the
case of fundamental strings 
which can end on D-branes (for a review see [\pol]).
All these arguments are of course related.
 The conservation
of brane charge method  has been tackled
by Townsend using a deformation 
argument which is known as
brane surgery [\pt]. This argument  is based
on the existence of Chern-Simons terms 
in supergravity theories,
the use of their field equations and 
certain assumptions about the
behaviour of the supergravity gauge potentials 
near the intersection of two or more branes. 
One of the advantages of 
brane surgery is
that it applies for branes in a curved background. 
In the context of effective brane 
worldvolume theories, the brane
intersections or boundaries are thought 
in a way similar to the
interactions of particles in 
standard quantum field theory.
So, 
the conservation of brane charges in this 
approach corresponds to
the conservation 
 of particle charges at the interactions 
vertices of the effective
brane theory.

From the (worldvolume) perspective of 
the branes involved in an intersection
or a boundary, the intersection or the 
boundary is described by a  defect.
 This defect can be
viewed as a soliton like object and its
charge can be measured on each brane involved 
in the configuration [\gppkt]. 
One expects that
the charges of the defects and the charges 
of the branes of the configuration are related
amongst themselves.  To derive
such relations, one first uses the fact that 
these defects  have a brane
interpretation so they are usually called
 worldvolume branes or brane 
worldvolume solitons. The brane surgery 
method of [\pt] then provides
a relation amongst the charges of the 
defects and the charges of the
branes of the configuration.

In this paper, we shall propose 
another way to describe the
brane charge conservation at the 
intersection or boundary of two or more branes.
Our construction is based on the Thom 
classes $\phi(E)$ of vector bundles
$E$ and their properties. In particular, 
 we view p-branes as (p+1)-dimensional
submanifolds $B_{p}$ in the spacetime $M$ which 
asymptotically behave as their
associated supergravity solutions. Their 
charge is then computed 
by integrating an appropriate form field 
strength at a sphere at infinity.
We shall see that this is the same as 
evaluating the Thom class $\phi(N_{(p)})$ of 
the normal bundle $N_{(p)}$ of $B_p$ in the
spacetime $M$; this has also been proposed by 
Witten in [\wittenb].  Next using the properties of 
Thom classes,  if  two
branes
$B_{p}$ and
$B_{q}$  intersect on $Z_k=B_p\cap B_q$, (${\rm dim} Z_k=k+1$),
we shall establish 
the relation
$$
Q_{(p)} Q_{(k\hra p)}=Q_{(q)} Q_{(k\hra q)}\ ,
\eqn\mainn
$$
where $Q_{(p)}$ is the charge of the p-brane, 
$Q_{(q)}$ is the charge of the q-brane, 
 $Q_{(k\hra p)}$ is the charge of the k-brane defect $Z_k$ on the
p-brane and $Q_{k\hra q}$ is the
charge of  $Z_k$
 on the q-brane. 
The charges $Q_{(k\hra p)}$ and $Q_{(k\hra q)}$ are associated
with the Thom classes of the normal bundles of $Z_k$ in $B_p$
and $B_q$, respectively.
The above relation among the various charges 
has two implications the following:

 \item{(i)}
The charges of the k-brane defect as measured on the
 p-brane and on the q-brane are proportional to the number of
q-branes and the number of p-branes involved in the configuration,
respectively.

\item {(ii)} There is a relation between the
units that one measures the charges of the 
defects with those of the p-and q-branes.

We shall  explore the relation \mainn\ amongst the charges 
for various boundaries and 
brane intersections in strings and 
M-theory. In particular, we shall find that 
the unit of the charge of the
boundary of a fundamental string ending on a 
D-brane is proportional
to the fundamental string tension.  Similarly, 
the unit of the charge of
a M-2-brane ending on a M-5-brane is proportional 
to the M-2-brane tension.
We shall also find that
in some cases the charges of the defects on 
intersecting branes
 are related to the 
Euler number of certain vector bundles. In addition, 
our method  can  be 
used to investigate charge
conservation for intersections that involve three or
more kinds of branes;  we
shall demonstrate this for a triple M-brane intersection.

This letter has been organized as 
follows: In section two, we use 
Thom classes to describe brane charge 
conservation for intersecting branes 
and derive the equation \mainn. In 
section three, we apply our 
formalism to investigate the 
conservation of charges in type II intersections.
In section four, we investigate 
charge conservation in M-brane intersections,
and in section five we present our conclusions and remark 
on the application of our results
in the context of supergravity 
solutions with the interpretation of
intersecting branes.

\chapter{Brane Charges and Thom Classes}
\section{The p-brane charge}

Spacetimes with a p-brane interpretation have an asymptotic
region which is isomorphic to $\bR^{(1,p)}\times \bR^{d-p-1}$, 
where $\bR^{(1,p)}$ 
are the worldvolume directions
 and $\bR^{d-p-1}$ are the 
transverse or normal directions of the p-brane [\gibbons].  
This is the so call spatial transverse
infinity which can be thought of as the spatial infinity
of the spacetime far away
from the location of the brane. Any p-brane in a d-dimensional
spacetime has an associated
$(d-p-2)$-form field strength $F$. The charge $Q_p$ per unit 
volume (in some frame) of the p-brane can be computed by evaluating
$F$ at a sphere $S^{d-p-2}\subset \bR^{d-p-1}$ at infinity as
$$
Q_p={1\over {\rm Vol}(S^{d-p-2})} \int_{S^{d-p-2}} F\ ,
\eqn\aone
$$
 where 
${\rm Vol}(S^{d-p-2})$ is the volume 
of the unit $S^{d-p-2}$ sphere.
Typically, the asymptotic behaviour of $F$ in some angular 
coordinates in $\bR^{d-p-1}$ is
$$
F= -{Q_p\over (d-p-3)} \star d{1\over  r^{d-p-3}}
\eqn\atwo
$$ 
as $r\rightarrow \infty$, where $r$ is 
the radius and the Hodge star
is that on $\bR^{d-p-1}$.

An alternative way to compute the charge 
of a brane is to observe that
$$
dF=Q_p\delta (r)
\eqn\athree
$$
where $\delta$ is  a  $(d-p-1)$-form with support at $r=0$.
So we can compute  $Q_p$ by
integrating $dF$ over $\bR^{d-p-1}$, i.e.
$$
Q_p=\int_{\bR^{d-p-1}} dF \ . 
\eqn\afout
$$
There is a  geometric way to view 
 this calculation. For this let us identify the
 p-brane as a submanifold 
$B_p$ of the spacetime $M$.  The normal bundle
$N_{(p)}$   of $B_p$ in $M$ is defined as
$$
TB_p\rightarrow TM|_{B_p}\rightarrow N_{(p)}\ .
\eqn\afive
$$
So at the transverse spatial infinity, 
$N_{(p)}= \bR^{(1,p)}\times \bR^{d-p-1}$. The
form $dF$ can be thought of as a $(d-p-1)$-form on $N_{(p)}$ 
which has support at the
zero section of $N_{(p)}$. In addition, the computation of 
charge of the p-brane above can be
thought of as the  integration of $dF$ along a 
fibre $\bR^{d-p-1}$ of $N_{(p)}$.

For every rank n vector bundle $E$ over a 
manifold $S$, the Thom class\foot{We assume 
that the manifolds and the
 vector bundles
involved are oriented.},  $\phi(E)$, of $E$  is a
(smooth) n-form in 
$\phi(E)$ which has the following properties:

\item{(i)} The integration of  $\phi(E)$ 
along any fiber of $E$ gives one.
\item{(ii)} $\phi(E)$ has support very
 close to the zero section of $E$.

We shall not give the details of the construction
 of the Thom class $\phi(E)$
of $E$. This is explained in [\bott]. The 
Thom class of the normal
bundle
$N_{(p)}$ of a submanifold
$B_p$ in
$M$ is the Poincar\'e dual of $B_p$.

Now we shall take $dF$ to be in the (cohomology) class of 
 $\phi(N_{(p)})$ of the normal bundle
$N_{(p)}$ of the p-brane $B_p$
 in the spacetime $M$. For this we  
appropriately rescale $\phi(N_{(p)})$ such
that  integration of $\phi(N_{(p)})$ over the
fibers of $N_{(p)}$ gives at transverse spatial 
infinity the charge $Q_p$ of the p-brane. 
The use of the Thom class $\phi(N_{(p)})$ instead of $dF$ 
to compute the charge\foot{One may wonder 
whether it is possible instead
of the Thom class of $N_p$ that is 
associated with $dF$ to use another 
class on $N_p$ that is
associated with
$F$  for the computation of the
charge everywhere on a p-brane.  However, no 
such class exists for $N_p$
 unless its Euler class
vanishes [\bott].}
of a p-brane has two advantages the following:

\item{(i)}The 
charge per unit volume of a p-brane 
can be computed not only at the
 transverse spatial infinity but at 
any point on the brane $B_p$.

\item{(ii)} The Thom class $\phi(N_{(p)})$ need not 
satisfy the supergravity field equations. 
Instead it is sufficient
  to assume
that there is a representative in the
class of $\phi(N_{(p)})$ that obeys the
supergravity field equations.

The latter point allows the
computation of the  charges at brane intersections
to be independent from the details of the dynamics.
In the remaining sections, we shall use the properties
of the Thom classes to investigate the charges
of worldvolume brane defects of intersecting branes.

\section{Intersecting branes}

The typical set up of an intersecting brane configuration
is that of a p-brane, $B_p$, and and 
a q-brane, $B_q$, intersecting
on a k-brane $Z_k=B_p\cap B_q$ in a spacetime $M$. 
Now since the defect $Z_k$ has  a brane 
interpretation as viewed from the
worldvolume perspective of both the p-brane and the q-brane,
we shall use the Thom classes of the normal 
bundles of $Z_k$ in $B_p$ and in
$B_q$ to compute its charges. For this, 
let $\phi(N_{(p)})$ and $\phi(N_{(q)})$
be the Thom classes of the normal 
bundles $N_{(p)}$ and $N_{(q)}$  of the p-brane and the q-brane
in the spacetime $M$,
respectively. Far away from the q-brane, 
the charge $Q_p$ of the p-brane
can be computed as in the previous section by integrating 
the Thom class $\phi(N_{(p)})$ along 
 a fibre of the normal bundle  
$N_{(p)}$.
Since this can be done at any point in $B_p$,
we can also evaluate the charge of the p-brane
at a point in the intersection $Z_k$.
However at the intersection $Z_k$ of the 
p-brane with the q-brane,  the normal bundle of the 
p-brane splits\foot{More precisely, we have
 $N_{(k\hra q)}\rightarrow N_{(p)}|_{Z_k}
\rightarrow N_{(p+q)}|_{Z_k}$.}  as
$$
N_{(p)}|_{Z_k}=N_{(p+q)}|_{Z_k}\oplus N_{(k\hra q)}
\eqn\nspit
$$
where $N_{(p+q)}=N_{B_p\cup B_q}$ is the normal 
bundle\foot{The fibre directions of this bundle are the
overall transverse directions of an intersecting brane
configuration in the 
terminology of [\gppkt].} of $B_p$ and $B_q$
in the spacetime $M$, 
and $N_{(k\hra q)}$ is the normal
 bundle of  $Z_k$ in $B_q$. 
This decomposition of $N_{(p)}|_{Z_k}$  can be seen by  
observing that   $Z_k$ is a 
submanifold of $B_q$ and so $N_{k\hra q}$ is a subbundle
of $N_{(p)}|_{Z_k}$.
Using the above splitting of $N_{(p)}$ 
and the properties of Thom classes,
we can write
$$
\phi(N_{(p)}|_{Z_k})=\phi(N_{(p+q)}|_{Z_k})
\wedge \phi(N_{(k\hra q)})\ .
\eqn\asix
$$   
In turn, this  implies that
$$
Q_{(p)}=Q_{(p+q)} Q_{(k\hra q)}\ ,
\eqn\aseven
$$
where $Q_{(p+q)}$ is interpreted as the
 charge of the \lq\lq whole configuration"
and $Q_{(k\hra q)}$ is interpreted the charge of the 
k-brane worldvolume defect from
the perspective of the q-brane.
Repeating the same argument for the q-brane, we find that
$$
Q_{(q)}=Q_{(p+q)} Q_{(k\hra p)}
\eqn\aeight
$$
where $Q_{(k\hra p)}$ is the charge of the 
k-brane worldvolume defect
from the perspective of the p-brane.
Eliminating $Q_{(p+q)}$,
we find
$$
Q_{(p)} Q_{(k\hra p)}=Q_{(q)} Q_{(k\hra q)}\ .
\eqn\keyy
$$ 
We have now derived the  equation \mainn\ of the introduction. 
The charge $Q_{(p)}$ can be written as
$$
Q_{(p)}=\mu_{(p)} n_p
\eqn\maone
$$
where $\mu_{(p)}$ is the unit of charge of the 
p-brane and $n_p\in \bZ$ is the 
number of p-branes of the configuration.
Similarly,  we can write $Q_{(q)}=\mu_{(q)} n_q$
for the q-brane. The equation \keyy\ is valid for
any number of p- and q-branes. This implies that
$$
\eqalign{
Q_{(k\hra p)}&=\mu_{(k\hra p)} n_q
\cr
Q_{(k\hra q)}&=\mu_{(k\hra q)} n_p\ ,}
\eqn\matwo
$$
and
$$
\mu_{(p)} \mu_{(k\hra p)}=\mu_{(q)} \mu_{(k\hra q)}\ ,
\eqn\ukeyy
$$
where $\mu_{(k\hra p)}$ and $\mu_{(k\hra q)}$ 
are the units of charges
of the k-brane worldvolume defects 
on the p-brane and the q-brane, respectively.
From these it is clear that the 
charge of the k-brane defect
on the p-brane is proportional 
to the number of q-branes, and similarly
the charge of the k-brane defect 
on the q-brane is proportional
to the number of p-branes of 
the configuration. This in fact
is what one naively expects in the 
context of defects which are 
associated with brane intersections.

The above computation can be 
easily extended to intersections
that involve more than two kinds 
of branes. We shall not attempt
to give the general analysis here. Instead, 
we shall find
the relations amongst the charges 
associated with a triple M-brane
intersection in an example below.

\chapter{Intersections in String Theory}

All the brane intersections in 
string theory can be derived
from brane intersections in 
M-theory using S- and T-dualities.
Nevertheless, some features of brane
intersections can be better understood 
in the context of string theory. We shall be  concerned
with the class of intersections 
which involves  two kinds of branes. 
 Our
intention is not to give a 
complete treatment of all possible
brane intersections in string theory but it is rather 
limited to present some examples
of the  method. From these it will become clear how one 
can investigate any brane intersection
in this way.

\section{Fundamental Strings Ending on D-branes}

One of the most well studied boundaries is 
that of a fundamental string ending
on a D-p-brane for $p\leq 6$. Using \keyy\  for 
$p=p$ and $q=1$, we find that
$$
Q_{(p)} Q_{(0\hra p)}= Q_{(0\hra 1)} Q_{(1)}\ .
\eqn\sedb
$$
For $p=0$, the Thom class of $N_{(0\hra 0)}$ 
can be taken to be the
zero form which implies that  
the charge $Q_{(0\hra 0)}$ vanishes. The 
equation \sedb\ remains
consistent provided $Q_{(0\hra 1)}=0$ and so 
one concludes that within this formalism 
 fundamental strings cannot 
end on a D-0-brane in agreement
with [\pt].  The equation \sedb\ for $p=1$ adapted for 
the units of charges reads\foot{We
shall add the
 subscripts $NS$ and $D$ with the obvious 
interpretation whenever it is
necessary 
  to avoid confusion.}
$$
\mu_{(1_D)} \mu_{(0\hra 1_D)}
= \mu_{(0\hra 1_{NS})} \mu_{(1_{NS})}\ .
\eqn\dfcon
$$
 This is the case
of a 
string junction  [\shrz]. It is natural to
identify  $\mu_{(0\hra 1_D)}$ with the unit of 
charge of the Born-Infeld
field of the D-string. To find the unit of this charge, we
first note that the D-string 
tension \foot{We use the string frame
 relation $T_{(p)}= g_s^a \mu_{(p)}$ between the
tension $T_{(p)}$ and the unit of charge $\mu_{(p)}$ of 
a p-brane, where $g_s$ is
the string coupling, $a=0$ for fundamental 
string, $a=-1$ for D-p-branes and $a=-2$
for NS-5-branes. In the 
conventions of [\schwarzb], 
$\mu_{(p)}= (2\pi)^{-p}(\alpha')^{-{p+1\over2}}$.} 
$T_{(1_D)}$ in the presence of
constant Born-Infeld field $F=F_{01}$  changes as
$$
T_{(1_D)}+{1\over2} {\mu_{(1_{NS})}\over g_s} (2\pi \alpha')^2 F^2\ .
\eqn\tenbi
$$
This follows from the Born-Infeld action of a D-string; 
a similar calculation has been
done in [\maldaa] for $\alpha'=1$.
On the other hand the tension of a state involving 
a D-string and a NS-string is 
$$
\mu_{(1_{NS})} \big(1+ g_s^{-2}\big)^{{1\over2}}=
({1\over g_s}+ {1\over2}g_s) \mu_{(1_{NS})}+
O(g_s^3)\ .
\eqn\tendn
$$
From the above two equations at small 
string coupling, we find that
$$
\mu_{(0\hra 1_D)}=F= g_s T_{(1_{NS})}\ .
$$
Substituting this into the equation \dfcon, we get
$$
\mu_{(0\hra 1_{NS})}=g_s T_{(1_{NS)}}\ .
\eqn\nsss
$$
So the defect on the fundamental string 
has unit charge proportional to
its tension. This may have been expected since the defect 
is a domain wall [\bergpap].

For the rest of the cases, we remark 
that the units of charges of 
 D-p-branes [\pol, \schwarz, \schwarzb, \itexas] are related
by the equation
$$
\mu_{(p)}={1\over 2\pi}\mu_{(1)} \mu_{(p-2)}\ .
\eqn\mone
$$
Using this equation and \ukeyy\ for $p=p$, 
$q=1$ and $k=0$, we find that
$$
\mu_{(0\hra 1)}={1\over 2\pi}\mu_{(p-2)} \mu_{(0\hra p)}\ .
\eqn\mtwo
$$
We  proceed by observing that the right hand 
side of \dfcon\ is T-duality invariant
provided that it is performed  in directions
orthogonal to the string. However, the left hand side
changes according to the familiar T-duality rules for D-branes
(for the T-duality rules of the 
defects see [\papastwo]). Moreover,  the
fact that in all cases the defect on the 
fundamental string is a domain wall suggests
that \nsss\ is valid for $0<p\leq 6$. Using these, we find
 that $\mu_{(0\hra p)}=2\pi g_s
\mu_{(1_{NS)}} \mu^{-1}_{(p-2)}$. This 
expression for $\mu_{(0\hra p)}$
can also be directly computed  in a way 
similar to that for $p=1$ above.

\section{Three-Brane Solitons in type IIA}

In type IIA, (i) two (non-parallel) 
NS-5-branes\foot{Cancellation of anomalies in the
effective theory of NS-5-branes 
requires that the Euler number of the
normal bundle of IIA NS-5-branes 
vanishes [\wittenb].} and (ii) a
NS-5-brane and a D-4-brane intersect on a 
three-brane soliton defect. We shall postpone the
investigation of (i) as well as that of the 
associated via T-duality intersection in type IIB for later.
This is  because these intersections
are similar to that of two M-5-branes 
intersecting on a three-brane. 
So all the details will follow
from the investigation of this intersection in M-theory.

Applying \keyy\ for the intersection of 
a NS-5-brane and a D-4-brane on a 
three-brane, 
we find
$$
Q_{(5)} Q_{(3\hra 5)}= Q_{(3\hra 4)} Q_{(4)}\ .
\eqn\sedbb
$$
The unit of charge  of the NS-5-brane and with 
that of D-4-brane are related [\schwarz, \itexas]
by 
$$
\mu_{(5)}={1\over 2\pi}\mu_{(0)} \mu_{(4)}\ .
\eqn\mfive
$$
Substituting this equation into \ukeyy\ for 
$p=5$, $q=4$ and $k=3$, we get
$$
{1\over 2\pi} \mu_{(0)} \mu_{(3\hra 5)}= \mu_{(3\hra 4)}\ .
\eqn\mtex
$$
If in addition we take $\mu_{(3\hra 4)}\sim \mu_{(4)}$, 
 because the 3-brane defect is a
domain wall in the D-4-brane, then 
$\mu_{(3\hra 5)}\sim \mu_{(4)} \mu^{-1}_{(0)}$.

\section{Two-Brane Solitons in type IIB}

In type IIB,  a D-3-brane intersects with a NS-5-brane 
on a 2-brane. Using \keyy\ for this intersection, 
we find
that 
$$
Q_{(5_{NS})} Q_{(2\hra 5_{NS})}= Q_{(2\hra 3_{D})} Q_{(3_{D})}\ .
\eqn\aaone
$$
The unit of charge of IIB  NS-5-brane and that of D-string
are related [\schwarz, \itexas] as 
$$
\mu_{(5_{NS})}={1\over 2\pi}\mu_{(1_D)} \mu_{(3_D)}\ .
\eqn\none
$$
Substituting this equation into \ukeyy\ for 
$p=5$, $q=3$ and $k=2$, we get
$$
{1\over 2\pi} \mu_{(1_D)} \mu_{(2\hra 5_{NS})}= \mu_{(2\hra 3_D)}
\eqn\mathree
$$
and thus we establish that the unit charge of  
2-brane defect on the NS-5-brane
and the unit of charge of 2-brane defect on 
the D-3-brane are related via the
unit of charge of D-string.

Finally in type IIB a 
NS-5-brane and a D-5-brane intersect on a
2-brane.
For this intersection,  \keyy\  becomes
$$
Q_{(5_{NS})} Q_{(2\hra 5_{NS})}= Q_{(2\hra 5_D)} Q_{(5_D)}\ .
\eqn\nsdfive
$$
The unit of charge of NS-5-brane and that of  D-5-brane
are related [\schwarz, \itexas] as
$$
\mu_{(5_{NS})}={1\over 2\pi}\mu_{(-1_D)} \mu_{(5_D)}\ ,
\eqn\bone
$$
which upon substitution in \ukeyy\ for 
$p=5$, $q=5$ and $k=2$ leads to
$$
{1\over 2\pi} \mu_{(-1_D)} \mu_{(2\hra 5_{NS})}= \mu_{(2\hra 5_D)}\ ,
\eqn\bsix
$$
where $\mu_{(-1_D)}$ is the 
unit of charge of  IIB instanton.

\chapter{M-Branes}

The are two  types of brane 
intersections [\gppkt, \tseytlinb] 
in M-theory\foot{In fact there is a third
non-standard type of string 
intersection that of two M-5-branes
intersecting at a string [\kastor].}. The first is
 that of a membrane  ending on 
a M-5-brane with defect a (self-dual) string.
The other is that two non-parallel M-5-branes 
intersecting  on a 3-brane.
The latter intersection is part of a more general intersection
rule which states that two non-parallel p-branes
intersect  on a (p-2)-brane. 
There is a large number of triple M-brane intersections.
Here we shall not do a systematic investigation of triple 
 intersections. Instead,
we shall examine the case of two (non-parallel) membranes
ending on a M-5-brane.

\section{Membranes ending on M-5-branes}

 The defect on a M-2-brane ending on a
 M-5-brane is a string. This defect from the perspective
of the M-5-brane is the self-dual string  
while from the perspective of
the M-2-brane is  a boundary.
Applying \keyy\ in this case for 
$p=5$, $q=2$ and $k=1$, we find
$$
Q_{(5)} Q_{(1\hra 5)}=Q_{(2)} Q_{(1\hra 2)}\ .
\eqn\cone
$$ 
Using the relation 
$\mu_{(5)}={1\over 2\pi}(\mu_{(2)})^2$ of the unit of charge
of the  M-5-brane\foot{The tension of M-branes in the
conventions of [\schwarzb] is 
$T_{(2)}=\mu_{(2)}=2\pi m_p^3$ and $T_{(5)}=\mu_{(5)}=2\pi
m_p^6$, where $m_p$ is the eleven-dimensional Plank mass.} and 
that of M-2-brane [\schwarz, \itexas], we can rewrite
\ukeyy\ adopted to this case as
$$
\mu_{(1\hra 2)}={1\over 2\pi} \mu_{(2)} \mu_{(1\hra 5)}\ .
\eqn\ctwo
$$
Therefore, the unit of charge $\mu_{(1\hra 2)}$ of the defect 
on the M-2-brane is proportional
to the unit of charge charge of the 
M-2-brane as it is expected for a domain wall. 
We remark that in this method the charge of the string defect
on the M-5-brane is not naturally associated with a self-dual
three-form as it may have been expected. 
However since the value of the
charge depends only the cohomology 
class of the form field strengths,
there may be a self-dual representative of this class.

\section{The 3-brane worldvolume soliton}

 The  relation \keyy\ of the various  charges in the
case of two non-parallel M-5-brane 
intersecting on a 3-brane defect is
$$
Q_{(5)} Q_{(3\hra 5)}=Q'_{(5)} Q'_{(3\hra 5)}\ ,
\eqn\cthree
$$
where $Q_{(5)}$ and $Q'_{(5)}$ are the charges of 
the two 5-branes and $Q_{(3\hra 5)}$ and 
$Q'_{(3\hra 5)}=Q_{(3\hra 5')}$
are the charges of the 3-brane 
defects, respectively.

To explore further the intersection 
of two M-5-branes on a
3-brane defect let us we assume that
 $Q_{(5)}=Q_{(5)}'$ and that the normal
bundle
$N_{(5)}$ of $B_5$ can be written as
$$
N_{(5)}=E\oplus H
\eqn\spit
$$  
where $E$ is a rank two bundle and $H$ is its 
compliment in $N_{(5)}$.
 Then  the intersection can be described 
as follows: we first identify   
 the M-5-brane $B_5$
as the image of  the zero section of $E$ while
we identify the second M-5-brane $B'_5$ as the image 
of as a generic section $s$ of $E$, i.e. $B'_5=s(B_5)$.
If $B_5$ and   $B'_5$ are in general position,
 then they intersect transversaly in $E$.
The 3-brane defect from the 
perspective of $B_5$ is the zero
locus $Z$ of the section $s$, i.e. $Z_3=B_5\cap s(B_5)$.
The normal bundle of $Z_3$ in $B_5$ is $E$ and 
in this case the Thom class of
$E$ can be identified with its Euler
 class $e(E)$. So the charge $Q_{(3\hra 5)}=Q_{(3\hra 5)}'$
can be identified with the Euler 
number of $E$ in some units.

\section{The 0-brane worldvolume soliton}

The method of relating the charges of
 various worldvolume defects associated
with brane intersections can be easily
 generalized to intersections
that involve three or more branes.
As an illustration we shall describe brane 
surgery for a configuration for which
the associated orthogonal
 M-brane intersection is
$$
\eqalign{
M5:& 0,1,2,3,4,5,-,-,-
\cr
M2:& 0,1,-,-,-,-,6,-,-
\cr
M2':&0,-,2,-,-,-,-,7,-\ .}
\eqn\caone
$$
We define $Z_1=B_2\cap B_5$, $Z_{1'}=
B_{2'}\cap B_5$ and $Z_0=Z_1\cap Z_{1'}=B_2\cap
B_{2'}\cap B_5$.
In such configuration the normal bundles of the M-branes
involved in the intersection split as follows:
$$
\eqalign{
N_{(2)}|_{Z_0}&=N_{(1\hra 5)}|_{Z_0}
\oplus N_{(1'\hra 2')}|_{Z_0}\oplus N_{(2+2'+5)}|_{Z_0}
\cr
N_{(2')}|_{Z_{0}}&=N_{(1'\hra 5)}|_{Z_0}
\oplus N_{(1\hra 2)}|_{Z_0}\oplus
N_{(2+2'+5)}|_{Z_0} 
\cr
N_{(2)}|_{Z_0}&=N_{(0\hra 2')}
\oplus N_{(1+1'\hra 5)}|_{Z_0}\oplus N_{(2+2'+5)}|_{Z_0}
\cr
N_{(2')}|_{Z_0}&=N_{(0\hra 2)}
\oplus N_{(1+1'\hra 5)}|_{Z_0}\oplus N_{(2+2'+5)}|_{Z_0}
\cr
N_{(5)}|_{Z_0}&= N_{(1\hra 2)}|_{Z_0}
\oplus N_{(1'\hra 2')}|_{Z_0} \oplus
N_{(2+2'+5)}|_{Z_0}\ .}
\eqn\catwo
$$
In addition we have
$$
\eqalign{
N_{(1\hra 5)}|_{Z_0}&=N_{(0\hra 1')}
\oplus N_{(1+1'\hra 5)}|_{Z_0}
\cr
N_{(1'\hra 5)}|_{Z_0}&=N_{(0\hra 1)}
\oplus N_{(1+1'\hra 5)}|_{Z_0}
\cr
N_{(0\hra 5)}|_{Z_0}&=
 N_{(0\hra 1)}\oplus N_{(0\hra 1')} 
\oplus N_{(1+1'\hra 5)}|_{Z_0}\ .}
\eqn\cathree
$$
The first two of above three 
decompositions can be understood by viewing
the strings $Z_1$ and $Z_{1'}$ as branes within the M-5-brane.
The above decompositions of the normal 
bundles lead to the following
relations amongst the brane charges: 
$$
\eqalign{
Q_{(2)}&=Q_{(1\hra 5)} Q_{(1'\hra 2')} Q_{(2+2'+5)}
\cr
Q_{(2')}&=Q_{(1'\hra 5)} Q_{(1\hra 2)} Q_{(2+2'+5)} 
\cr
Q_{(2)}&=Q_{(0\hra 2')} Q_{(1+1'\hra 5)} Q_{(2+2'+5)}
\cr
Q_{(2')}&=Q_{(0\hra 2)} Q_{(1+1'\hra 5)} Q_{(2+2'+5)}
\cr
Q_{(5)}&= Q_{(1\hra 2)} Q_{(1'\hra 2')}  Q_{(2+2'+5)}\ ,}
\eqn\cafive
$$
and
$$
\eqalign{
Q_{(1\hra 5)}&=Q_{(0\hra 1')} Q_{(1+1'\hra 5)}
\cr
Q_{(1'\hra 5)}&=Q_{(0\hra 1)} Q_{(1+1'\hra 5)}
\cr
Q_{(0\hra 5)}&= Q_{(0\hra 1)} 
Q_{(0\hra 1')} Q_{(1+1'\hra 5)}\ .}
\eqn\casix
$$
Using the above relations, we find
$$
\eqalign{
Q_{(5)} Q_{(0\hra 5)}&= Q_{(0\hra 2)} Q_{(2)}
\cr
Q_{(5)} Q_{(0\hra 5)}&= Q_{(0\hra 2')} Q_{(2')}\ ,}
\eqn\casev
$$
which leads to
$$
Q_{(5)} Q_{(0\hra 5)}=\big(Q_{(0\hra 2)} Q_{(0\hra 2')}
Q_{(2)} Q_{(2')}\big)^{{1\over2}}\ .
\eqn\bbone
$$
This relates the charges of the defect as measured
on membranes and M-5-brane with the charges of membranes
and M-5-brane involved in the configuration.
The above computation can be easily extended 
to many other triple brane 
intersections.

\chapter{Concluding Remarks}

We have proposed a method to investigate
 charge conservation at the
intersection of two or more branes based 
on Thom classes. We have found
that this has led to a relation between the charges of 
the branes involved in the
intersection
 and those 
 of the associated worldvolume defects. 
We have then explored
these relations for various brane 
intersections in strings and M-theory.

Some brane 
intersections preserve a
proportion of spacetime supersymmetry. 
This can be incorporated
in our brane surgery construction 
by imposing additional restrictions
 on the spacetime and the submanifolds associated
with the various branes. For example, 
one can introduce a supersymmetry
projection operator at every point in 
the submanifold associated with a brane in a way
similar to that of [\kal] and then ask whether
there are killing spinors that satisfy all these projections.
However as we have seen brane charge conservation can be
investigated without the additional restriction of supersymmetry

It is natural to ask whether 
there are solutions in the
literature that satisfy all the 
requirements necessary to establish the
above relations amongst the charges of the branes
and those of the defects. 
This is related to the
question of localization of the 
brane intersection solutions. It
has been observed in the beginning of 
construction of supergravity solutions
with the interpretation as brane
intersections [\gppkt, \gp] that they 
are smeared along their 
relative transverse directions
and that their asymptotic behaviour does not have 
the desired power decay law
with respect to some radial coordinate. The latter
leads to problems for calculating the charges at infinity
of the associated branes. Thus such 
intersections are geometrical
and there is no defect on either brane involved in the 
intersection. Subsequent improvements in the 
solutions by adding different
harmonic functions for each intersecting  brane 
[\tseytlinb] have not resolve the problem.
More recently solutions 
[\yang, \youm, \ansar, \loewy] have been found 
using the so called generalized
harmonic function equations first proposed in [\papgib].
Such solutions exhibit a partial 
localization at the intersection.
However, their associated 
form-field strengths do not have
the desirable asymptotic behaviour
required for the evaluation of brane charges.
Because of these and other arguments, it was 
suggested in [\peet] that for many
intersecting brane configurations 
there are not exit solutions which
are completely localized.
However it may simply be that supergravity solutions
that exhibit  charge conservation at brane intersections
are not simply constructed from harmonic functions and
their (straightforward) generalizations.
So it appears that new methods should be developed to 
solve the supergravity
field equations like those for
 BPS monopoles in gauge theories.

\vskip 1cm
\noindent{\bf Acknowledgments:}    I would 
like to thank  
K.S. Stelle for correspondence, and Paul Townsend
for many helpful discussions and suggestions. I am
supported by a University Research Fellowship from the Royal
Society.

\refout

\bye